\documentclass[bibyear]{aa}
\usepackage{graphicx}
\usepackage{txfonts}
\newcommand{\RSTAR}{\mbox{$R_{\star}$}}
\newcommand{\MSTAR}{\mbox{$M_{\star}$}}
\newcommand{\TEFF}{\mbox{$T_{\rm eff}$}}

\newcommand{\MSOL}{\mbox{$M_{\sun}$}}
\newcommand{\LSOL}{\mbox{$L_{\sun}$}}
\newcommand{\MBOL}{\mbox{$M_{\rm BOL}$}}
\newcommand{\MSOLPERYR}{\mbox{$M_{\sun}$~yr$^{-1}$}}
\newcommand{\micron}{\mbox{$\mu$m}}
\newcommand{\KMS}{\mbox{km s$^{-1}$}}
\newcommand{\HOH}{\mbox{H$_2$O}}
\newcommand{\PERSQCM}{\mbox{cm$^{-2}$}}
\newcommand{\VMICRO}{\mbox{$\varv_{\rm micro}$}}
\newcommand{\LOGG}{\mbox{$\log \varg$}}
\newcommand{\alfcenA}{\mbox{$\alpha$~Cen~A}}
\newcommand{\swvir}{\mbox{SW~Vir}}
\newcommand{\bkvir}{\mbox{BK~Vir}}
\begin{document}

\title{
Spatially resolving the atmosphere of the non-Mira-type AGB star SW~Vir 
in near-infrared molecular and atomic lines with VLTI/AMBER
\thanks{
Based on AMBER observations made with the Very Large Telescope 
Interferometer of the European Southern Observatory. 
Program ID: 092.D-0461(A)}
\fnmsep
\thanks{
Observed data shown in Fig.~\ref{obsres} 
are available in electronic form
at the CDS via anonymous ftp to cdsarc.u-strasbg.fr (130.79.128.5)
or via http://cdsweb.u-strasbg.fr/cgi-bin/qcat?J/A+A/}
}

\author{K.~Ohnaka\inst{1}, 
M.~Hadjara\inst{1,2}, 
and M.~Y.~L.~Maluenda Berna\inst{1}
}

\offprints{K.~Ohnaka}

\institute{
Instituto de Astronom\'ia, Universidad Ca\'olica del Norte, Avenida Angamos
0610, Antofagasta, Chile\\
\email{k1.ohnaka@gmail.com}
\and
Centre de Recherche en Astronomie, Astrophysique et G\'eophysique (CRAAG), 
Route de l'Observatoire, B.P. 63, Bouzareah, 16340, Alger, Alg\'erie
}

\date{Received / Accepted }

\abstract
{}
{
We present a near-infrared spectro-interferometric observation of the 
non-Mira-type, semiregular asymptotic giant branch star SW~Vir. 
Our aim is to probe the physical properties of the outer atmosphere 
with spatially resolved data in individual molecular and atomic lines. 
}
{
We observed SW~Vir in the spectral window between 2.28 and 2.31~\micron\ 
with the near-infrared interferometric instrument AMBER 
at ESO's Very Large Telescope Interferometer (VLTI). 
}
{
Thanks to AMBER's high spatial resolution and 
high spectral resolution of 12\,000, the atmosphere of 
SW~Vir has been spatially resolved not only in strong CO 
first overtone lines but also in weak molecular and atomic lines of 
\HOH, CN, HF, Ti, Fe, Mg, and Ca. 
While the uniform-disk diameter of the star is 16.23$\pm$0.20~mas 
in the continuum, it increases up to 22--24~mas in the CO lines. 
Comparison with the MARCS photospheric models reveals that the star 
appears larger than predicted by the hydrostatic models 
not only in the CO lines but also even in the weak 
molecular and atomic lines. 
We found that 
this is primarily due to the \HOH\ lines (but also possibly 
due to the HF and Ti lines) originating in the extended outer atmosphere. 
Although the \HOH\ lines manifest themselves very little in the spatially 
unresolved spectrum, the individual rovibrational \HOH\ 
lines from the outer atmosphere 
can be identified in the spectro-interferometric data. 
Our modeling suggests an \HOH\ column density of 
$10^{19}$--$10^{20}$~\PERSQCM\ in the outer atmosphere extending out to 
$\sim$2~\RSTAR. 
}
{
Our study has revealed that the effects of the nonphotospheric outer 
atmosphere are present in the spectro-interferometric data not only in 
the strong CO first overtone lines but also in the weak molecular and 
atomic lines. 
Therefore, analyses of spatially unresolved spectra, such as for example 
analyses of the chemical composition, should be carried out 
with care even if the lines appear to be weak. 
}


\titlerunning{
Spatially resolving the atmosphere of \swvir\ in near-infrared molecular 
and atomic lines
}
\authorrunning{K. Ohnaka}
\maketitle

\begin{table*}
\caption {
Summary of the VLTI/AMBER observations of \swvir\ and the calibrator \alfcenA. 
}
\begin{center}

\begin{tabular}{l c c c c r l }\hline
\# & $t_{\rm obs}$ & $B_{\rm p}$ & PA     & Seeing   & $\tau_0$ &
${\rm DIT}\times{\rm N}_{\rm f}\times{\rm N}_{\rm exp}$ \\ 
   & (UTC)       & (m)       & (\degr) & (\arcsec)       &  (ms)    &  (ms)  \\
   &             &B2-C1/C1-D0/B2-D0 & B2-C1/C1-D0/B2-D0 & &         &        \\
\hline
\multicolumn{7}{c}{\swvir : 2014 February 11 (UTC)}\\
\hline
1 & 06:32:29 & 9.8/19.6/29.4 & 14/14/14    & 0.78 & 5.9 & $120\times500\times5$ \\
\hline
\multicolumn{7}{c}{\alfcenA: 2014 February 11 (UTC)}\\
\hline
C1 & 05:02:03 & 10.07/20.17/30.23 & 15/15/15 & 1.02& 4.8 &$120\times500\times5$\\
C2 & 05:36:38 & 10.14/20.32/30.46 & 9/9/9 & 0.97& 5.0&$120\times500\times5$\\
C3 & 06:51:22 & 10.19/20.40/30.59 & 3/3/3 & 0.94& 4.9 &$120\times500\times5$\\
C4 & 07:26:01 & 10.15/20.33/30.48 & 9/9/9 & 0.82& 5.5 &$120\times500\times5$\\
C5 & 08:02:11 & 10.07/20.18/30.25 & 14/14/14 & 1.26& 3.5&$120\times500\times5$\\
\hline
\label{obslog}
\vspace*{-7mm}

\end{tabular}
\end{center}
\tablefoot{
$B_{\rm p}$: Projected baseline length.  PA: position angle of the baseline 
vector projected onto the sky. 
DIT: detector integration time.  $N_{\rm f}$: number of frames in each 
exposure.  $N_{\rm exp}$: number of exposures. 
The seeing and the coherence time ($\tau_0$) were measured in the visible. 
}
\end{table*}

\section{Introduction}
\label{sect_intro}

Mass loss in late evolutionary stages of low- and intermediate-mass stars 
is important not only for the evolution of stars themselves but also 
for the chemical enrichment of the interstellar medium. 
In particular, the stars in the asymptotic giant branch (AGB) 
experience mass loss with mass-loss rates of $\sim \!\! 10^{-8}$~\MSOLPERYR\ 
up to $\sim \!\! 10^{-4}$~\MSOLPERYR. 
Nevertheless, the mass-loss mechanism in AGB stars is not yet fully 
understood. The levitation of the atmosphere by the large-amplitude pulsation 
and the radiation pressure on dust grains are a viable mechanism 
to drive the mass loss in Mira-type AGB stars (e.g., H\"ofner \& Olofsson 
\cite{hoefner18}). 
However, non-Mira-type AGB stars with 
semiregular and irregular variability with much smaller variability amplitudes 
also show noticeable mass loss. At the moment, it is not clear how the 
mass loss is driven in these non-Mira-type AGB stars. 
The fact that the number of non-Mira-type AGB stars is comparable to that of 
Mira stars as mentioned by Ohnaka et al. (\cite{ohnaka12}) highlights the 
importance of improving our understanding of the mass-loss mechanism in these 
non-Mira-type AGB stars. 

The acceleration of stellar winds in AGB stars is considered to take place 
within several stellar radii. 
High spatial resolution achieved by infrared long-baseline interferometry 
provides an excellent opportunity to spatially resolve this key region 
and improve our understanding of its physical properties. 
Taking advantage of the high spatial resolution and high spectral 
resolution (up to $\lambda/\Delta \lambda$=12\,000) 
of the near-infrared interferometric instrument AMBER 
(Petrov et al. \cite{petrov07}) at 
ESO's Very Large Telescope Interferometer (VLTI), we have been studying 
the outer atmosphere of K--M giants. The VLTI/AMBER 
observations in the individual CO first overtone lines near 2.3~\micron\ 
have allowed us to spatially resolve the molecular outer 
atmosphere---so-called MOLsphere as coined by Tsuji (\cite{tsuji00})---and 
derive its physical properties such as radius, temperature, and CO column 
density in three K and M giants, Arcturus ($\alpha$~Boo, K1.5III), 
Aldebaran ($\alpha$~Tau, K5III), and BK~Vir (M7III) 
(Ohnaka \& Morales Mar\'in \cite{ohnaka18}; Ohnaka \cite{ohnaka13a}; 
Ohnaka et al. \cite{ohnaka12}). 
These studies show that the MOLsphere is extending out to 
2--3 \RSTAR\ even in K giants, which cannot be explained by the current 
hydrostatic photospheric models. 
The presence of the MOLsphere in K--M giants 
has also been revealed by infrared interferometric 
observations with lower spectral resolutions 
across the molecular bands (instead of individual lines) of CO, \HOH, 
and TiO 
(e.g., Quirrenbach et al \cite{quirrenbach93}; 
Mennesson et al. \cite{mennesson02}; 
Sacuto et al. \cite{sacuto08}, \cite{sacuto13}; 
Mart\'i-Vidal et al. \cite{marti-vidal11}; 
Arroyo-Torres et al. \cite{arroyo-torres14}). 

The high spectral resolution of VLTI/AMBER of 12\,000 also allows us to probe 
the atmospheric structure not only with the strong CO lines but also 
with weaker molecular and atomic lines. 
They are considered to form in deeper 
photospheric layers, and therefore high-spectral- and high-spatial-resolution 
observations in the weak lines provide us with an opportunity to 
probe the atmospheric layers different from those studied with the CO lines. 
In this paper we present the results of a spatially resolved observation of 
the non-Mira-type AGB star SW~Vir in the strong CO first overtone lines 
as well as in weaker molecular and atomic lines near 2.3~\micron. 

The M7III star SW~Vir is one of the nearby red giants at a distance of 
$143^{+19}_{-15}$~pc (based on a parallax of $6.99\pm 0.83$~mas, 
van Leeuwen \cite{vanleeuwen07}). 
It is a semiregular variable with a period of 150~days, 
classified as 
SRB in the General Catalogue of Variable Stars 
(Samus et al. \cite{samus17}). 
The light curve compiled by the American Association of Variable Star 
Observers (AAVSO) shows a variability amplitude of $\sim$1~mag 
(from minimum to maximum) in the visible. 
The mass-loss rate and the expansion velocity of the stellar wind 
of \swvir\ are estimated to be 
$(1.7-9.8) \times 10^{-7}$~\MSOLPERYR\ and 7.5--8.5~\KMS, respectively 
(Knapp et al. \cite{knapp98}; 
Gonz\'alez Delgado et al. \cite{gonzalez_delgado03}; 
Winters et al. \cite{winters03}). 
Despite its small variability amplitude, the mass-loss rate of \swvir\ 
is comparable to that of some of the optically bright Miras 
with much larger variability amplitudes (see the samples of 
Knapp et al. \cite{knapp98}, Gonz\'alez Delgado et al. 
\cite{gonzalez_delgado03}, and \mbox{Winters} et al. \cite{winters03}). 
The spatially unresolved spectroscopic studies of infrared molecular lines 
such as CO, \HOH, OH, SiO, and CO$_2$ in \swvir\ 
reveal that they cannot be entirely 
reproduced by the current photospheric models, suggesting the presence of 
the MOLsphere (Tsuji \cite{tsuji88}, \cite{tsuji08}; 
Tsuji et al. \cite{tsuji94}, \cite{tsuji97}). 
Also, Mennesson et al. (\cite{mennesson02}) found that the $L^{\prime}$-band 
(3.79~\micron) diameter of \swvir\ is 1.4 times larger than that in the 
$K^{\prime}$ band (2.13~\micron), which can be explained by molecular 
emission (presumably \HOH) from the MOLsphere. 

We describe our AMBER observation and data reduction in 
Sect.~\ref{sect_obs} and the observational results in Sect.~\ref{sect_res}. 
The comparison with the current photospheric models and our modeling of 
the observed data are presented in Sect.~\ref{sect_modeling}, followed 
by concluding remarks in Sect.~\ref{sect_concl}.

\begin{figure*}
\sidecaption
\resizebox{\hsize}{!}{\rotatebox{0}{\includegraphics{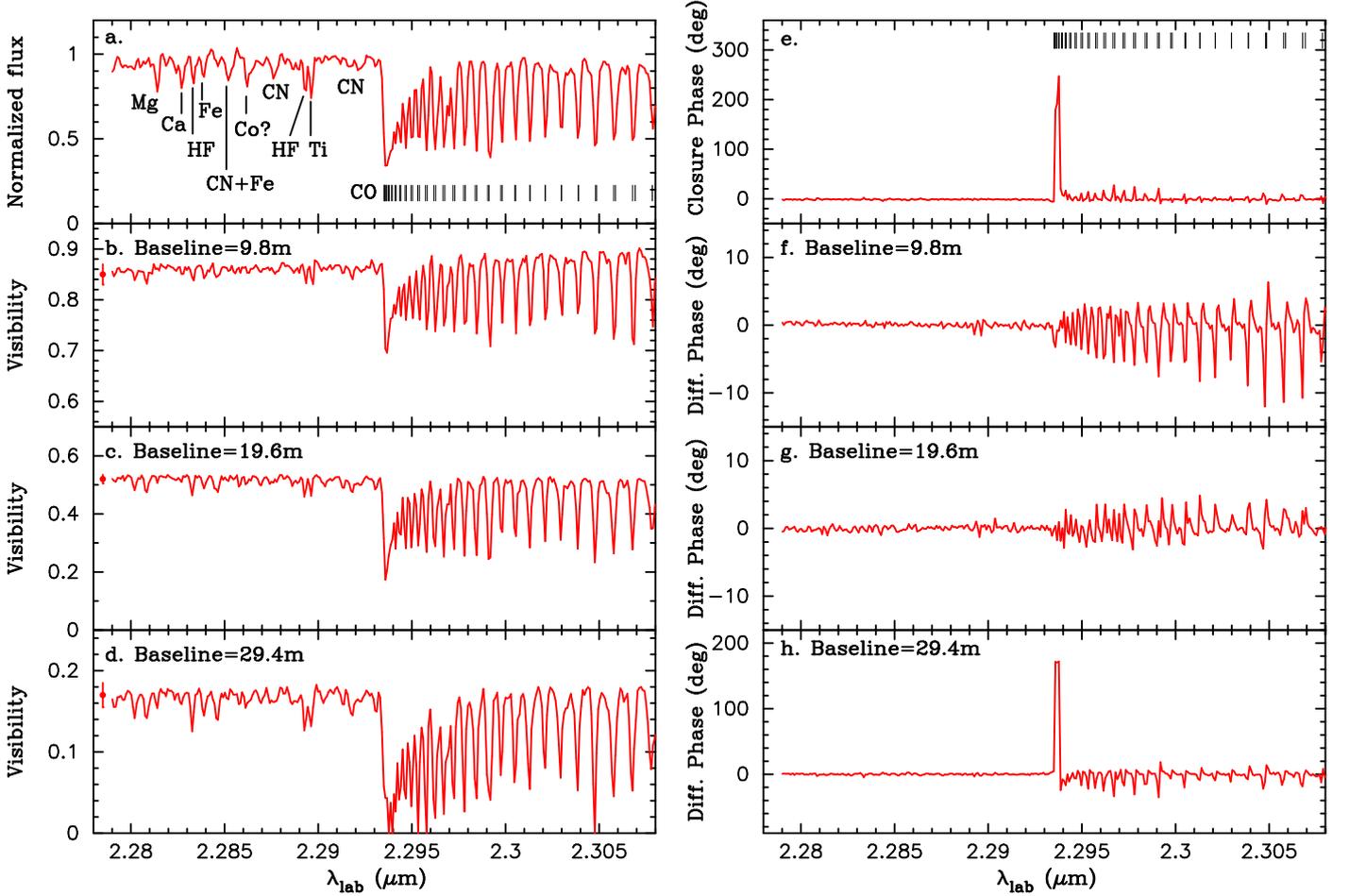}}}
\caption{
VLTI/AMBER observation of \swvir\ in the CO first overtone lines as well 
as in weak atomic and molecular lines. 
{\bf a:} Observed spectrum. The identification of some of the lines is shown. 
{\bf b--d:} Visibilities observed at the baselines of 9.8, 19.6, and 29.4~m, 
respectively. The typical errors in the visibilities are shown on the left. 
{\bf e:} Observed closure phase. The typical errors are $\sim$5\degr. 
The error bars are not shown because they would appear too small to recognize 
in the plot. The ticks show the positions of the CO lines. 
{\bf f--h:} Differential phases observed at the baselines of 
9.8, 19.6, and 29.4~m, respectively. 
The typical errors at each baseline are as follows: 
9.8~m: 0.5\degr--1\degr\ throughout the spectral window. 
19.6~m: 0.5\degr--1\degr\ in the continuum and in the CO lines and 
$\sim$3\degr\ at the CO band head. 
29.4~m: 1\degr--2\degr\ in the continuum, 10\degr--20\degr\ in the CO lines, 
and $\sim$40\degr\ at the CO band head. 
}
\label{obsres}
\end{figure*}

\section{Observation and data reduction}
\label{sect_obs}

Our VLTI/AMBER observation of SW~Vir took place on 2014 February 11 (UTC) using 
the Auxiliary Telescope (AT) configuration of B2-C1-D0, which resulted in 
projected baseline lengths of 9.8, 19.6, and 29.4~m. 
As in our previous studies, we observed 
the spectral region between 2.28 and 2.31~\micron\ with a spectral resolution 
of 12\,000 without the VLTI fringe tracker FINITO. 
A summary of our AMBER observations of SW~Vir and its calibrator \alfcenA\ 
(G2V) is given in Table~\ref{obslog}. The calibrator \alfcenA\ was observed 
five times throughout the night not only for the observation of \swvir\ 
but also for other targets in the program. 

The data were reduced with amdlib ver.~3.0.7\footnote{http://www.jmmc.fr/data\_processing\_amber.htm}, which is 
based on the P2VM algorithm (Tatulli et al. \cite{tatulli07}; 
Chelli et al. \cite{chelli09}). 
We adopted a uniform-disk diameter of $8.314 \pm 0.016$~mas for the 
calibrator \alfcenA\ (Kervella et al. \cite{kervella03}) for the 
calculation of the transfer function. 
The interferometric data of \swvir\ were calibrated using the transfer 
function values measured before and after \swvir. However, using 
only two measurements before and after \swvir\ can underestimate the 
errors in the transfer function. Therefore, we adopted the standard deviation 
($1\sigma$) of the transfer function values measured throughout the night 
as the errors in the transfer function. 
The atmospheric conditions (seeing and coherence time) were stable throughout 
the night, which resulted in a very stable transfer function during the 
night (see Appendix of Ohnaka \& Morales Mar\'in \cite{ohnaka18}). 
We compared the calibrated interferometric observables---squared visibility 
amplitude, 
closure phase (CP), and differential phase (DP)---derived by taking the best 
20\% and 80\% of the frames in terms of the fringe S/N. 
We took the squared visibility amplitude obtained with the best 20\% of the 
frames, because the errors are smaller than with the best 80\%. For the CP 
and DPs, we took the results with the best 80\% because of their smaller
errors. 

The wavelength calibration was done using the telluric lines observed 
in the spectrum of \alfcenA. The uncertainty in the wavelength calibration 
is $2.1 \times 10^{-5}$~\micron, which corresponds to 2.7~\KMS. The observed 
wavelength scale was converted to the laboratory frame, using the 
heliocentric velocity of $-15.0$~\KMS\ of \swvir\ 
(Gontcharov \cite{gontcharov06}) and the correction for the motion of 
the observer in the direction of the observation computed with the 
IRAF\footnote{IRAF is distributed by the National Optical Astronomy
  Observatory, which is operated by the Association of Universities for
  Research in Astronomy (AURA) under a cooperative agreement with the National
  Science Foundation.} 
task \verb|rvcorrect|. 
The spectroscopic calibration to correct for the telluric lines and 
instrumental effects was done using the observed spectrum of \alfcenA, 
as described in Ohnaka et al. (\cite{ohnaka13b}).

\section{Results}
\label{sect_res}

The calibrated visibilities, CP, and DPs are shown in Fig.~\ref{obsres}. 
The observed spectrum, plotted in Fig.~\ref{obsres}a, shows that 
the individual CO lines are clearly resolved. 
The visibilities measured at three baselines 
show noticeable decrease in the individual CO lines, which means that the 
star appears larger in the CO lines than in the continuum. 

Furthermore, the observed spectrum shows some weak lines shortward of 
the CO band head at 2.294~\micron. We identified them to be the lines of 
neutral atoms (Mg, Ca, Ti, and Fe) and molecules (CN and HF) by comparing 
with the high-resolution spectrum of the M8 giant RX~Boo presented by 
Wallace \& Hinkle (\cite{wallace96}). 
The observed visibilities, particularly the one obtained at the longest 
baseline (Fig.~\ref{obsres}d), show signatures of these weak lines. 
It is worth noting that the visibility dips do not necessarily 
correspond to the lines identified in the spectrum. For example, the
visibility dips at 2.2802 and 2.2808~\micron\ 
(Figs.~\ref{obsres}a--\ref{obsres}c) do not correspond to any 
clear lines but only to a weak trough in the spectrum (Fig.~\ref{obsres}a). 
This puzzling point will be analyzed in detail in 
Sect.~\ref{subsect_model_weak}.

\begin{figure}
\sidecaption
\resizebox{\hsize}{!}{\rotatebox{0}{\includegraphics{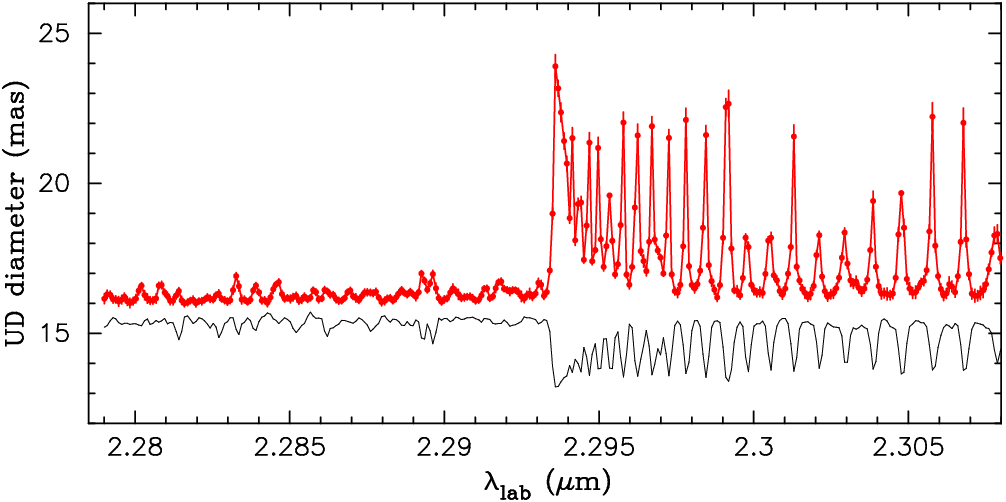}}}
\caption{
Uniform-disk diameter of \swvir\ (red line with the error bars). 
The scaled observed spectrum is shown with the black solid line. 
}
\label{udfit}
\end{figure}

Figure~\ref{udfit} shows the uniform-disk diameter obtained by fitting the 
visibilities obtained at three baselines at each wavelength channel. 
The uniform-disk diameter is $16.23\pm 0.20$~mas in the continuum, 
while it increases up to 22--24~mas (36--48\% with respect to the continuum 
diameter) in the CO band head and strong CO lines. 
The uniform-disk diameter in the continuum is in agreement with 
the previous measurements around 2.2~\micron\ 
($16.81\pm0.12$~mas, Perrin et al. 
\cite{perrin98}; $16.24\pm0.06$~mas, Mennesson et al. \cite{mennesson02}; 
$15.9\pm 0.6$~mas, Mondal \& Chandrasekhar \cite{mondal05}). 
The uniform-disk diameter also increases by $\sim$5\% in the weak lines 
shortward of the CO band head, reflecting the visibility dips described 
above. 

\begin{figure}
\sidecaption
\resizebox{\hsize}{!}{\rotatebox{0}{\includegraphics{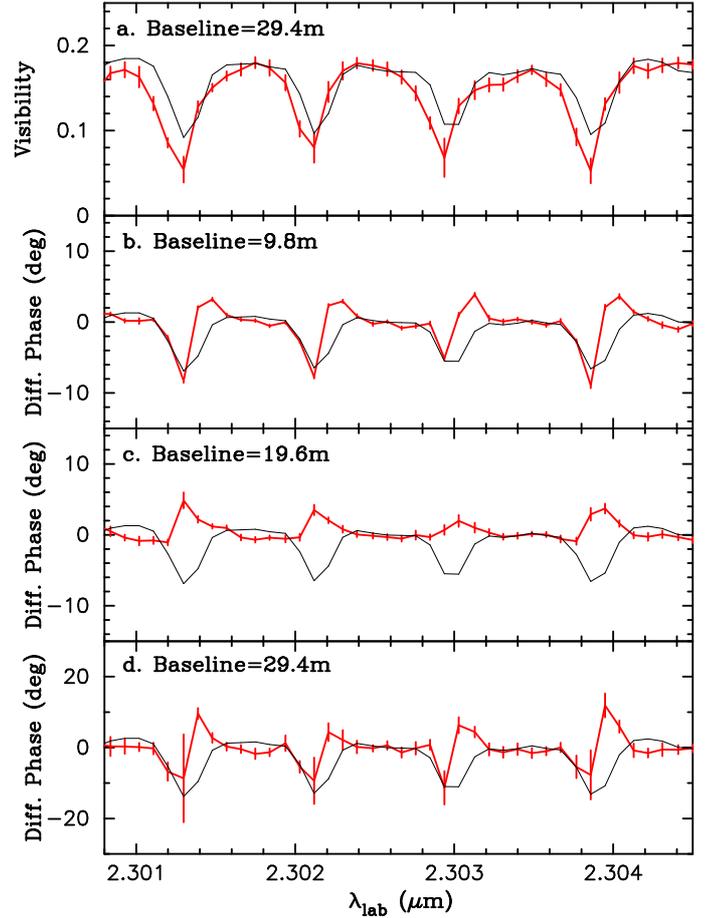}}}
\caption{
Enlarged view of the visibility obtained at the longest baseline of 29.4~m 
(panel {\bf a}) and the differential phases observed at three baselines 
(panels {\bf b}--{\bf d}). In each panel, the interferometric observable 
(visibility or differential phase) is shown with the red line with the 
error bars, while the black line represents the scaled observed spectrum. 
}
\label{obsresCO}
\end{figure}

The CP and DPs observed in the individual CO lines show nonzero or 
non-180\degr\ values (Figs.~\ref{obsres}e--\ref{obsres}h), 
which indicate asymmetry in the outer atmosphere where the CO lines form. 
Furthermore, as shown in Fig.~\ref{obsresCO}, the visibility observed 
at the longest baseline of 29.4~m and the DPs measured at all three baselines 
are asymmetric with respect to the center of the line profile. 
The visibility minima at the 29.4~m baseline (Fig.~\ref{obsresCO}a) are 
shifted to the blue wing of the lines. The DPs measured at the 9.8 and 29.4~m 
show minima in the blue wing and maxima in the red wing. The DP measured at 
the 19.6~m baseline is also asymmetric with respect to the line center but 
the asymmetry is much less pronounced compared to the other two baselines. 

The asymmetry detected in the visibility and DPs 
means that the star appears differently across the line profiles. 
The analysis of the AMBER data of the red supergiants Betelgeuse and 
Antares shows that inhomogeneous velocity fields can make the star appear 
differently across the CO line profile (Ohnaka et al. \cite{ohnaka09}, 
\cite{ohnaka11}, \cite{ohnaka13b}). The velocity-field map of Antares 
reconstructed from the AMBER data indeed reveals inhomogeneous, turbulent 
motions of large gas clumps (Ohnaka et al. \cite{ohnaka17}). 
Therefore, the asymmetry in the visibility and DPs seen in \swvir\ 
suggests that there might be inhomogeneous motions in the CO-line-forming upper 
photosphere and outer atmosphere.

\section{Modeling of the AMBER data}
\label{sect_modeling}

\subsection{Determination of basic stellar parameters}
\label{subsect_param}

It is necessary to determine basic stellar parameters of \swvir\ so that 
we can specify MARCS photospheric models (\mbox{Gustafsson} et al. 
\cite{gustafsson08}) appropriate for this star. 
We collected photometric data from the visible to the mid-infrared: 
broadband photometry from the $U$ to $K$ band 
(Ammons et al. \cite{ammons06}; 
Ducati \cite{ducati02}; 
Cutri et al. \cite{cutri03}; Hall \cite{hall74}; 
Kharchenko \& Roeser \cite{kharchenko09}; 
Mermilliod et al. \cite{mermilliod87}; 
Zacharias et al. \cite{zacharias05})
and the spectrophotometric data of up to 45~\micron\ 
obtained with the Short-Wavelength Spectrometer 
(SWS) onboard the Infrared Space Observatory (ISO) 
(Sloan et al. \cite{sloan03})\footnote{
https://users.physics.unc.edu/\~{}gcsloan/library/swsatlas/atlas.html}. 
The data were de-reddened with $A_V = 0.077$, which was obtained using the 
interstellar extinction map presented by Arenou et al. (\cite{arenou92}). 

The bolometric flux computed by integrating these (spectro)photometric data 
is $7.0 \times 10^{-9}$~W~m$^{-2}$.  We estimated the uncertainty in the 
bolometric flux to be 6\% based on the variations in the collected 
photometric data. 
Combined with the distance of $143^{+19}_{-15}$~pc, 
this results in a luminosity of $4500 \pm 1100$~\LSOL\ (\MBOL\ = $-4.4$), 
which agrees very well with the value obtained by Tsuji (\cite{tsuji08}). 
Using the continuum angular diameter of $16.23 \pm 0.20$~mas, we derived 
the effective temperature to be $2990 \pm 50$~K. This effective temperature 
also agrees with the $2886\pm 100$~K obtained by Tsuji (\cite{tsuji08}) 
using the infrared flux method. 
We estimated a stellar mass to be 1--1.25~\MSOL\ based on a comparison with the 
theoretical evolutionary tracks from Marigo et al. (\cite{marigo13}) 
(see also Fig.~13 in Rau et al. \cite{rau17}) and also from Lagarde et al. 
(\cite{lagarde12}). 
The surface gravity (\LOGG) is then estimated to be $-0.3 \pm 0.1$.

In addition to these basic stellar parameters, 
it is necessary to specify the micro-turbulent velocity (\VMICRO) and chemical 
composition to select a MARCS model appropriate for \swvir. 
We adopted a micro-turbulent velocity of 4~\KMS\ based on the analysis 
of high-resolution spectra of the CO lines (Tsuji \cite{tsuji08}). 
The C, N, and O abundances of \swvir\ indicate the mixing of the 
CN-cycled material (Tsuji \cite{tsuji08}). 
The sum of the C, N, O abundances of \swvir\ derived by 
Tsuji (\cite{tsuji08}) suggests a solar or slightly subsolar metallicity. 

The MARCS model with the parameters closest to the observationally 
derived values has \TEFF\ = 3000~K, \LOGG\ = 0.0, \MSTAR\ = 1~\MSOL, 
\VMICRO\ = 2~\KMS, 
[Fe/H] = 0.0, and the moderately CN-cycled composition.

\subsection{CO first overtone lines}
\label{subsect_model_cofov}

\begin{figure}
\sidecaption
\resizebox{\hsize}{!}{\rotatebox{0}{\includegraphics{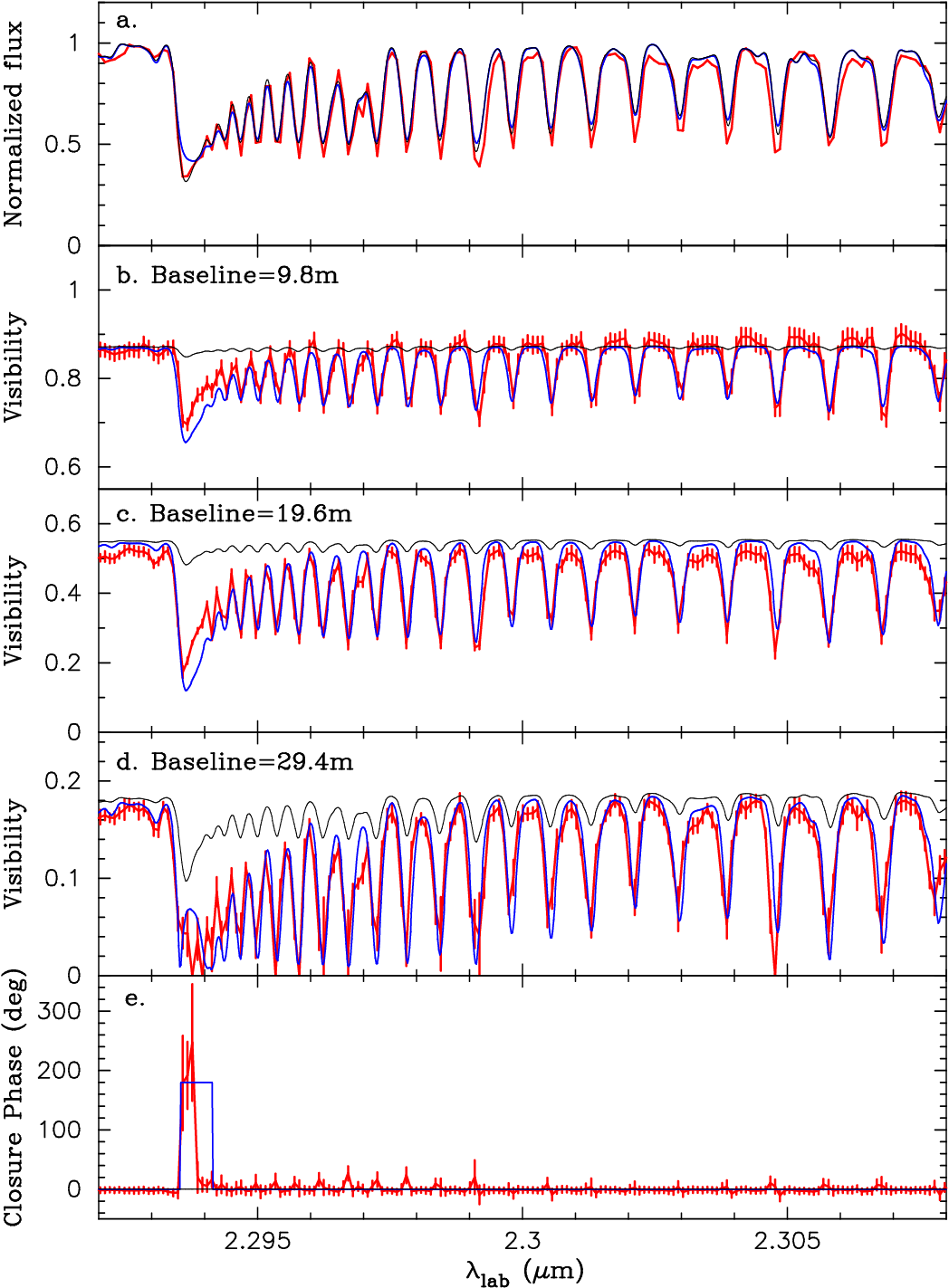}}}
\caption{
Comparison of the AMBER data obtained in the CO first overtone lines 
with the MARCS-only model and MARCS+MOLsphere model for \swvir. 
{\bf a:} Spectrum.
{\bf b--d:} Visibilities observed at the baselines of 9.8, 19.6, and 29.4~m, 
respectively. 
{\bf e:} Closure phase. 
In each panel, the red thick line with the error bars shows the observed 
data. The thin solid black line represents the MARCS-only model, while the 
thin solid blue line represents the MARCS+MOLsphere model. 
In panel {\bf a}, the black and blue solid lines are almost entirely 
overlapping. 
}
\label{marcs_wme_cofov}
\end{figure}

Our previous AMBER observations of the non-Mira-type AGB star BK~Vir, 
very similar to \swvir, show the presence of an extended component 
not accounted for by the hydrostatic photospheric models 
(Ohnaka et al. \cite{ohnaka12}). 
However, 
Arroyo-Torres et al. (\cite{arroyo-torres14}) show that 
the AMBER observations of five red giant stars 
(four stars probably on the AGB and one star on the red giant branch) 
across the CO bands at 2.3--2.45~\micron\ can be explained by hydrostatic 
models. 
Therefore, 
we first compared the AMBER data observed in the CO first overtone lines 
with the MARCS model selected for \swvir. 
The spectrum and visibility were calculated as described in 
Ohnaka (\cite{ohnaka13a}), using the CO line list of Goorvitch 
(\cite{goorvitch94}). We also included lines of \HOH, HF, CN, and atomic 
lines, which we discuss in more detail in Sect.~\ref{subsect_model_weak}, 
although their contribution is minor compared to the CO lines in the 
spectral region longward of 2.2936~\micron. 
The spectrum and visibility predicted by the MARCS model are shown by 
the thin black lines in Fig.~\ref{marcs_wme_cofov}. Although the observed 
spectrum is well reproduced (the thin black line is almost entirely 
overlapping with 
the blue solid line that we discuss below), the visibilities predicted by 
the MARCS model are too high compared to the observed data. 
This is what was observed during our previous studies of K--M giants 
(Ohnaka \& Morales Mar\'in \cite{ohnaka18}; Ohnaka \cite{ohnaka13a}; 
Ohnaka et al. \cite{ohnaka12}). 
As Ohnaka \& Morales Mar\'in (\cite{ohnaka18}) discuss, the negative 
detection of an extended outer atmosphere by Arroyo-Torres et al. 
(\cite{arroyo-torres14}) may be due to the lower spectral resolution of 
1500 used in their AMBER observations. 

We applied the semi-empirical models that consist of the MARCS photospheric 
model and additional two CO layers and searched for the best-fit parameters 
for the CO layers, as described in Ohnaka et al. (\cite{ohnaka12}).  
Figure~\ref{marcs_wme_cofov} shows a comparison of the 
observed data with the best-fit MARCS+2-layer model (blue solid lines). 
This model is characterized by the inner CO layer located at 1.3~\RSTAR\ 
with a CO column density of $2\times 10^{22}$~\PERSQCM\ and a temperature 
of 2000~K and the outer CO layer located at 2.0~\RSTAR\ with a CO column 
density of $10^{20}$~\PERSQCM\ and a temperature of 1700~K. 
The model can reasonably reproduce not only the observed spectrum but also the 
visibilities observed on all three baselines. 
Also, the model predicts the CP at the band head at 2.2936~\micron\ 
to be 180\degr, which roughly agrees with the observed data. 
This is because the MOLsphere makes the object appear large enough that 
the longest baseline of 29.4~m corresponds to the second visibility lobe, 
where the phase is 180\degr. 

\begin{figure*}
\sidecaption
\rotatebox{0}{\includegraphics[width=12cm]{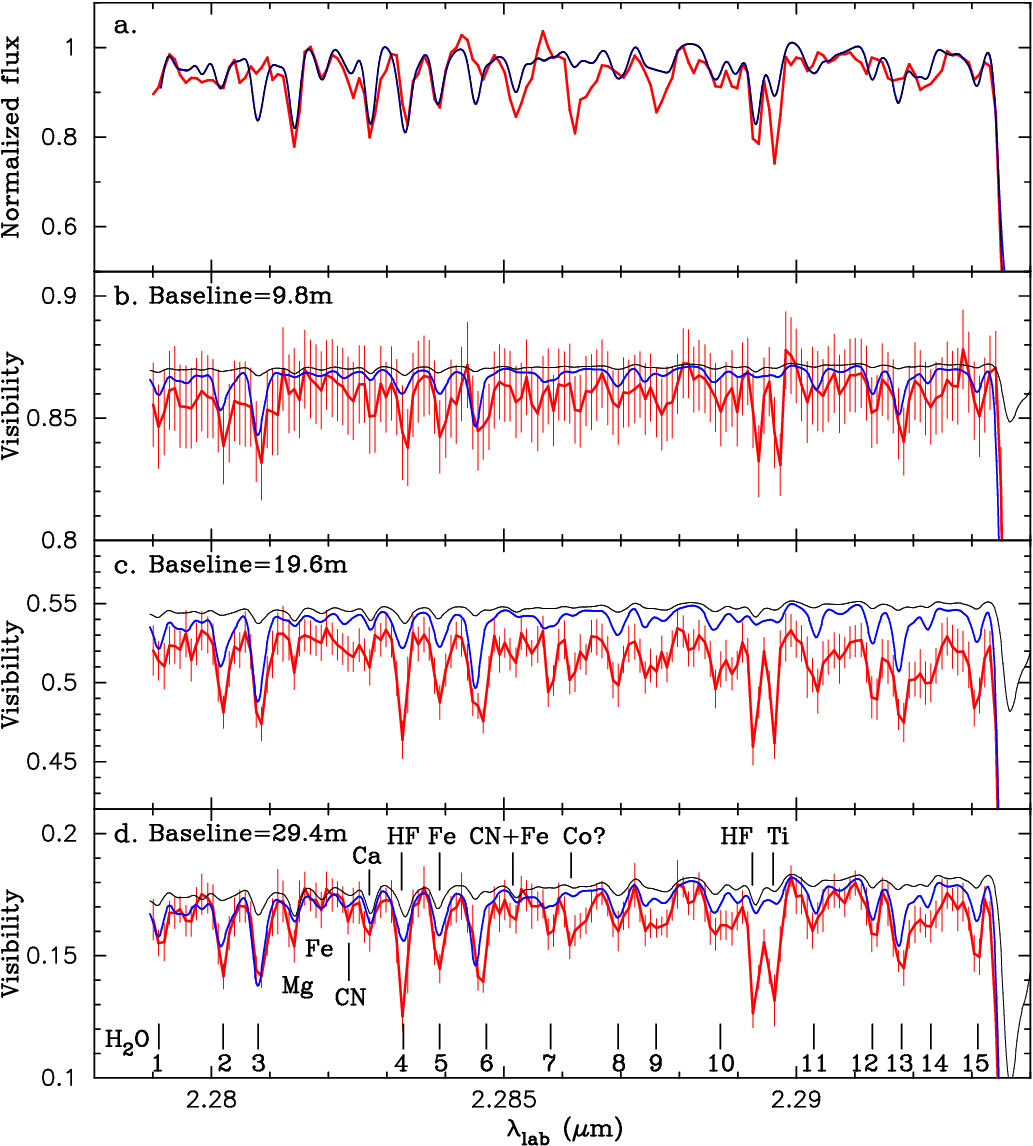}}
\caption{
Comparison of the AMBER data obtained in the weak atomic and molecular lines 
with the MARCS-only model and MARCS+MOLsphere model for \swvir. 
{\bf a:} Spectrum.
{\bf b--d:} Visibilities observed at the baselines of 9.8, 19.6, and 29.4~m, 
respectively. 
In each panel, the red thick line with the error bars shows the observed 
data. The thin solid black line represents the MARCS-only model, while the 
thin solid blue line represents the MARCS+MOLsphere model. 
In panel {\bf a}, the black and blue solid lines are almost entirely 
overlapping. 
In panel {\bf d}, the ticks in the bottom mark 
the \HOH\ lines responsible for the visibility dips. 
Their identification is listed in Table~\ref{water_line_id}. 
The ticks indicate the position of the \HOH\ features seen in the observed 
spectrum. 
As the table shows, they consist of more than one \HOH\ line in some cases. 
}
\label{marcs_wme_weak}
\end{figure*}

\begin{figure}
\sidecaption
\resizebox{\hsize}{!}{\rotatebox{0}{\includegraphics{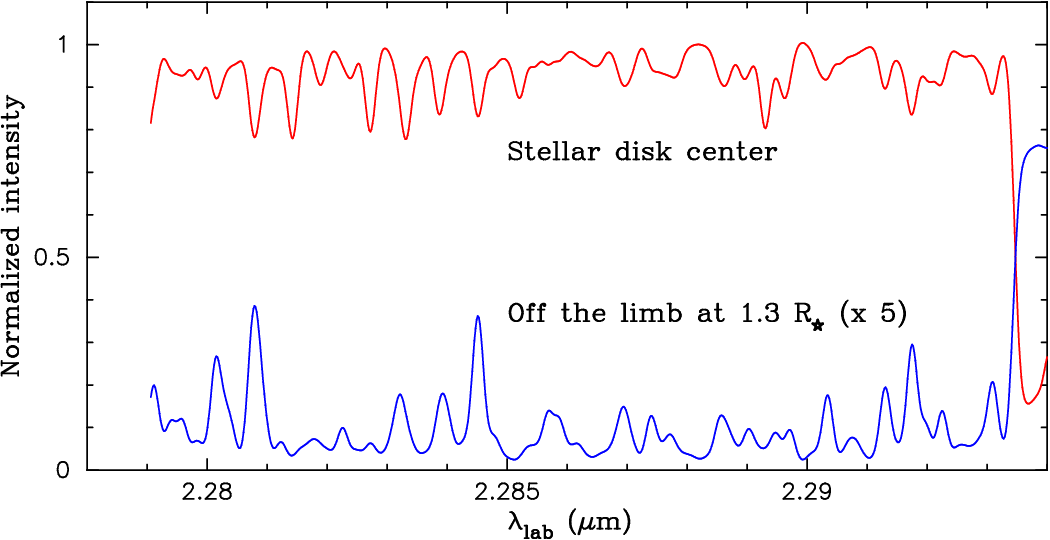}}}
\caption{
Spatially resolved synthetic spectra of the \HOH\ lines 
at the stellar disk center (red line) 
and off the limb at 1.3~\RSTAR\ (blue line). 
This latter spectrum is scaled by a factor of five for an easier comparison. 
Both spectra were obtained from the MARCS+MOLsphere model with \HOH\ described 
in Sect.~\ref{subsect_model_weak} and shown in Fig.~\ref{marcs_wme_weak}. 
}
\label{intens_sp}
\end{figure}

\begin{table*}
\begin{center}
\caption {\HOH\ lines identified in the observed visibilities of \swvir. 
The ID numbers correspond to those marked in Fig.~\ref{marcs_wme_weak}d. 
}

\begin{tabular}{rccccccc}\hline
\rule{0pt}{1.0\normalbaselineskip}
\#ID & Wavenumber & $\log \varg f$ &$E_{\rm exc}^{\prime\prime}$&
$(\varv_{1}^{\prime}\varv_{2}^{\prime}\varv_{3}^{\prime})$ & 
$(J^{\prime}K_{a}^{\prime}K_{c}^{\prime})$ & 
$(\varv_{1}^{\prime\prime}\varv_{2}^{\prime\prime}\varv_{3}^{\prime\prime})$ & 
$(J^{\prime\prime}K_{a}^{\prime\prime}K_{c}^{\prime\prime})$ \\
 & (cm$^{-1}$) &                 & (cm$^{-1}$) & & & & \\
\hline
\rule{0pt}{1.0\normalbaselineskip}
1 & 4387.663 & $-5.05$ &  5680.651 & (011) & (18,6,12) & (010) & (17,4,13)\\
2 & 4385.683 & $-5.70$ &  2745.973 & (001) & (15,6,10) & (000) & (14,4,11)\\
  & 4385.387 & $-6.38$ &  2748.068 & (001) & (14,8,7)  & (000) & (13,6,8)\\
3 & 4384.531 & $-5.72$ &  2918.212 & (001) & (15,7,9)  & (000) & (14,5,10)\\
  & 4384.385 & $-5.72$ &  3080.123 & (001) & (16,5,11) & (000) & (15,3,12)\\
4 & 4379.783 & $-5.16$ &  5835.418 & (011) & (18,7,11) & (010) & (17,5,12)\\
5 & 4378.527 & $-6.59$ &  2918.212 & (020) & (15,8,7)  & (000) & (14,5,10)\\
  & 4378.453 & $-6.06$ &  3386.457 & (011) & (13,4,10) & (010) & (12,2,11)\\
  & 4378.246 & $-5.76$ &  4728.315 & (011) & (16,5,11) & (010) & (15,3,12)\\
6 & 4377.360 & $-5.88$ &  2756.395 & (001) & (14,8,6)  & (000) & (13,6,7)\\
  & 4377.242 & $-5.33$ &  4017.868 & (100) & (18,7,12) & (000) & (17,4,13)\\
7 & 4375.077 & $-6.25$ &  3939.886 & (110) & (15,3,12) & (010) & (14,2,13)\\
  & 4375.062 & $-5.88$ &  4506.801 & (110) & (16,5,12) & (010) & (15,2,13)\\
  & 4374.736 & $-5.70$ &  5835.418 & (030) & (18,8,11) & (010) & (17,5,12)\\
  & 4374.685 & $-6.54$ &  2915.939 & (011) & (12,2,10) & (010) & (11,0,11)\\
8 & 4372.638 & $-6.47$ &  2321.863 & (001) & (12,9,3)  & (000) & (11,7,4)\\
9 & 4371.797 & $-6.38$ &  2248.002 & (001) & (14,5,10) & (000) & (13,3,11)\\
  & 4371.162 & $-6.66$ &  3770.864 & (011) & (11,9,3)  & (010) & (10,7,4)\\
10& 4369.591 & $-6.25$ &  3811.957 & (100) & (18,5,13) & (000) & (17,4,14)\\
  & 4368.680 & $-6.18$ &  3833.288 & (011) & (12,8,4)  & (010) & (11,6,5)\\
11& 4366.149 & $-6.06$ &  2433.768 & (001) & (13,8,6)  & (000) & (12,6,7)\\
12& 4364.332 & $-6.22$ &  1774.679 & (001) & (13,4,10) & (000) & (12,2,11)\\
13& 4363.477 & $-5.09$ &  4174.306 & (001) & (18,7,11) & (000) & (17,5,12)\\
14& 4362.541 & $-6.52$ &  2437.475 & (001) & (13,8,5)  & (000) & (12,6,6)\\
  & 4362.515 & $-6.19$ &  3437.206 & (100) & (17,6,12) & (000) & (16,3,13)\\
15& 4360.920 & $-5.32$ &  5015.841 & (011) & (16,7,9)  & (010) & (15,5,10)\\
  & 4360.901 & $-5.71$ &  5193.936 & (021) & (13,5,9)  & (020) & (12,3,10)\\
\hline
\label{water_line_id}
\vspace*{-7mm}

\end{tabular}
\end{center}
\end{table*}

\subsection{Weak molecular and atomic lines}
\label{subsect_model_weak}

As Fig.~\ref{obsres}a shows, the weak lines of CN, HF, Mg, Ti, Ca, and Fe are 
present shortward of the CO band head. 
In general, weak lines form in deep layers, where the atmospheric structure 
is considered to be reasonably described by hydrostatic photospheric 
models. Therefore, weak molecular and atomic lines are often used for the 
determination of chemical composition in red giant stars in combination 
with hydrostatic photospheric models 
(e.g., 
Abia et al. \cite{abia17}; Ga\l an et al. \cite{galan17}; 
Rich et al. \cite{rich17}; Do et al. \cite{do18}; 
D'Orazi et al. \cite{dorazi18}; \mbox{Thorsbro} et al. \cite{thorsbro18}). 
Therefore, it is meaningful to see whether the AMBER 
data in these weak lines can be explained by the photosphere alone without 
the MOLsphere. 
We computed the synthetic visibilities and spectrum from the MARCS model, 
including the lines of $^{12}$C$^{14}$N and $^{13}$C$^{14}$N 
using the line list of 
Sneden et
al. (\cite{sneden14})\footnote{https://www.as.utexas.edu/\~{}chris/lab.html}, 
assuming $^{12}$C/$^{13}$C = 22 (Tsuji \cite{tsuji08}). 
The line data of HF were taken from Jorissen et al. (\cite{jorissen92}). 
The atomic line 
data were taken from Kurucz \& Bell (\cite{kurucz95})\footnote{https://www.cfa.harvard.edu/amp/ampdata/kurucz23/sekur.html}. 
We also included $^{1}$H$_{2}$$^{16}$O lines using the list published 
by Barber et al. 
(\cite{barber06})\footnote{http://exomol.com/data/molecules/H2O/1H2-16O/BT2/}. 

Figure~\ref{marcs_wme_weak} shows a comparison of the AMBER data with 
the MARCS model (thin black solid lines). 
The figure shows that the observed spectrum of 
the weak lines is fairly reproduced. 
However, the model predicts the visibility dips 
observed in most of the weak lines to be too small compared to the observed 
data. It is particularly clearly seen in the data obtained at the longest 
baseline (Fig.~\ref{marcs_wme_weak}d). 
Furthermore, as mentioned in 
Sect.~\ref{sect_res}, the observed visibility dips do not necessarily 
correspond to the features seen in the spectrum. These results suggest that 
some molecular or atomic species in the extended MOLsphere may be 
making the star appear larger also in the weak lines than in the continuum, 
as in the case of the CO lines. 

We found out that the positions of the visibility dips correspond well to 
the positions of \HOH\ lines. 
To check whether or not the \HOH\ lines in the extended outer atmosphere can 
explain the observed visibility dips, 
we added \HOH\ in the best-fit MARCS+2-layer 
model obtained in Sect.~\ref{subsect_model_cofov}. 
Figure~\ref{marcs_wme_weak} shows the spectrum 
and visibilities predicted by the best-fit model with \HOH\ in the MOLsphere 
(blue solid lines). 
The model is characterized with an \HOH\ column density of
$10^{20}$~\PERSQCM\ in the inner layer and $2\times 10^{19}$~\PERSQCM\ 
in the outer 
layer (other parameters are set the same as the best-fit model with CO alone). 
Obviously, the visibility dips are much better reproduced not only at the 
longest baseline but also at the shortest and middle baselines 
(the offset between the data and the model seen at the middle baseline 
is due to the uncertainty in the absolute visibility calibration). 
The identification of the individual rovibrational 
\HOH\ lines responsible for the visibility dips 
is listed in Table.~\ref{water_line_id}. 

There are still several visibility dips 
that cannot be reproduced even with 
the MOLsphere model with \HOH. The visibility dips in the 
HF lines at 2.2832~\micron\ and 2.2893~\micron\ as well as the Ti line at 
2.2896~\micron\ are much deeper than predicted by the model. 
This indicates the presence of HF and Ti in the MOLsphere. 
Nevertheless, it is difficult to confirm it with only one or two lines, and 
therefore, we refrain from modeling these HF and Ti lines. 
The visibility dip at 2.2862~\micron\ coincides with a Co line. 
However, it is very weak in the synthetic spectrum predicted by the MARCS 
model, and the \HOH\ or CN lines cannot explain it either. 
While it is possible that it indicates the presence of Co in the MOLsphere, 
it may be affected by the blend of (an) unidentified line(s). 
Therefore, this feature remains unidentified at the moment.

The presence of \HOH\ in the MOLsphere in non-Mira-type 
K--M giants is known 
from the previous spectroscopic analyses (e.g., Tsuji et al. \cite{tsuji97}; 
Tsuji \cite{tsuji01}, \cite{tsuji08}) and is also revealed by 
the spectro-interferometric observations at 2--2.45~\micron\ with VLTI/AMBER 
by Mart\'i-Vidal et al. (\cite{marti-vidal11}) and 
Arroyo-Torres et al. (\cite{arroyo-torres14}). 
The resolution of these studies of 1500 was sufficient to 
resolve the \HOH\ band but not sufficient to resolve individual \HOH\ lines. 
The present work is the first study to spatially resolve individual 
\HOH\ lines originating in the MOLsphere. 

It should be noted that the \HOH\ lines originating in the MOLsphere 
manifest themselves very little in the spectrum (as can be seen when comparing the spectra 
predicted by the MARCS-only model and MARCS+MOLsphere model in 
Fig.~\ref{marcs_wme_weak}a), 
although their effects on the visibilities can be clearly recognized. 
This is explained as follows. As Fig.~\ref{intens_sp} shows, on the one hand, 
the spatially resolved synthetic spectrum predicted at the center of the 
stellar disk (red line) 
shows the individual \HOH\ lines in absorption, because we see 
the cooler MOLsphere in front of the warmer photosphere. 
On the other hand, the spatially resolved spectrum predicted off the limb
of the star (blue line) 
shows the \HOH\ lines in emission, because we do not see the warmer 
radiation from the photosphere in this case (Kirchhoff's law). 
The spatially unresolved spectrum is the sum of the spatially resolved 
spectra over the stellar disk and the extended atmosphere. 
Therefore, the \HOH\ lines in absorption are filled in by the emission 
from the same \HOH\ lines originating in the extended atmosphere, 
resulting in the spatially unresolved spectrum without conspicuous 
signatures of the \HOH\ lines.

\section{Concluding remarks}
\label{sect_concl}

We have spatially resolved the non-Mira-type AGB star \swvir\ 
not only in the 
CO first overtone lines but also in the weak atomic and molecular lines, 
taking advantage of the high spatial and spectral resolution of 
VLTI/AMBER. 
The AMBER data reveal that the star appears larger than 
the MARCS photospheric model predicts, not only in the CO first overtone 
lines but also even in many of the weak atomic and molecular lines 
observed between 2.28 and 2.294~\micron. 
This new observational result about the weak lines can be explained primarily 
by the \HOH\ lines originating in the extended outer atmosphere. 
Our modeling suggests the presence of 
\HOH\ with column densities of $10^{19}$--$10^{20}$~\PERSQCM\ in the 
outer atmosphere extending to $\sim$2~\RSTAR. 
Therefore, our spatially resolved observation 
of the individual \HOH\ lines confirms the presence of \HOH\ in the extended 
molecular atmosphere. 
An important point is that while the effects of the \HOH\ lines are clearly 
visible in the spectro-interferometric data, the \HOH\ lines themselves 
manifest very little in the spatially unresolved spectrum 
owing to the filling-in effect. 
This means that analyses of spatially unresolved spectra of red giants 
should be carried out with care (e.g., determination of chemical 
composition), even if the lines appear to be weak. 

The recent theoretical modeling of dynamical atmospheres including 
pulsation and/or convection covers the stellar parameters of relatively 
cool non-Mira-type AGB stars with $\TEFF \la 3200$, including \swvir\ 
(Bladh et al. \cite{bladh15}; Freytag et al. \cite{freytag17}). 
It would be of great interest to see whether or not the present 
spectro-interferometric AMBER data can be explained by these dynamical 
models. 

The MOLsphere parameters of \swvir\ derived from the CO first overtone lines 
are similar to those derived for another  
M7 giant, \bkvir, by Ohnaka et al. (\cite{ohnaka12}). 
The basic stellar parameters of \bkvir\ are very similar to those of \swvir. 
Therefore, it is not very surprising that the MOLsphere 
of two stars is similar, 
which in turn implies that the MOLsphere parameters might depend on 
basic stellar parameters. 
This point will be studied in more detail in a future paper based on the 
AMBER data that we have obtained for a small sample of red giant stars 
with spectral types ranging from K1.5 to M8, for most of which 
no theoretical dynamical models exist yet.

\begin{acknowledgement}
We thank the ESO VLTI team for supporting our AMBER observation. 
K.~O. acknowledges the support of the 
Comisi\'on Nacional de Investigaci\'on Cient\'ifica y Tecnol\'ogica
(CONICYT) through the FONDECYT Regular grant 1180066. 
The present work was also financed by the ALMA-CONICYT Fund, allocated to 
the project No. 31150002. 
This research made use of the \mbox{SIMBAD} database, 
operated at the CDS, Strasbourg, France, 
and NSO/Kitt Peak FTS data on the Earth's telluric features 
produced by NSF/NOAO.
We acknowledge with thanks the variable star observations from the AAVSO 
International Database contributed by observers worldwide and used in 
this research.

\end{acknowledgement}

\end{document}